\documentclass[twocolumn,english,amssymb,prl,showpacs]{revtex4}
\usepackage[T1]{fontenc}
\usepackage[latin9]{inputenc}
\usepackage{amsmath}
\usepackage{amssymb}
\usepackage{graphicx}
\usepackage{esint}

\makeatletter
\@ifundefined{textcolor}{}
{%
 \definecolor{BLACK}{gray}{0}
 \definecolor{WHITE}{gray}{1}
 \definecolor{RED}{rgb}{1,0,0}
 \definecolor{GREEN}{rgb}{0,1,0}
 \definecolor{BLUE}{rgb}{0,0,1}
 \definecolor{CYAN}{cmyk}{1,0,0,0}
 \definecolor{MAGENTA}{cmyk}{0,1,0,0}
 \definecolor{YELLOW}{cmyk}{0,0,1,0}
 }

\makeatother

\usepackage{babel}
\begin{document}

\title{Charge localization and dynamical spin locking in double quantum
dots driven by ac magnetic fields}

\author{\'{A}lvaro G\'{o}mez-Le\'{o}n}

\affiliation{Instituto de Ciencia de Materiales de Madrid (CSIC), Cantoblanco,
E-28049 Madrid, Spain}

\author{Gloria Platero}

\affiliation{Instituto de Ciencia de Materiales de Madrid (CSIC), Cantoblanco,
E-28049 Madrid, Spain}

\date{\today}
\begin{abstract}
In this work we investigate electron localization and dynamical spin
locking induced by ac magnetic fields in double quantum dots. We demonstrate
that by tuning the ac magnetic fields parameters, i.e., the field
intensity, frequency and the phase difference between the fields within
each dot, coherent destruction of tunneling (and thus charge localization)
can be achieved. We show that in contrast with ac electric fields,
ac magnetic fields are able to induce spin locking, i.e., to freeze
the electronic spin, at certain field parameters. We show how the
symmetry of the Hamiltonian determines the quasienergy spectrum which
presents degeneracies at certain field parameters, and how it is reflected
in the charge and spin dynamics. 
\end{abstract}
\maketitle

\paragraph*{Introduction:}

Quantum coherent effects in mesoscopic systems, such as quantum dots,
are a subject of great current interest, both from the theoretical
point of view, and because of a growing number of possible applications.
Experimental successes in detecting Rabi oscillations driven by electric
ac-fields in single quantum dots (QD's)\cite{Oosterkamp1998} have
spurred interest in the use of intense ac-fields to coherently manipulate
the time development of electronic states\cite{Cole2001}. An exciting
possibility is to make use of the phenomenon of coherent destruction
of tunneling (CDT)\cite{Grossmann1992,Grossmann1991,GrifoniM.1998,Frasca2005,Barata2000,Lignier2007,Della2007},
in which the tunneling dynamics of a quantum system becomes suppressed
at certain values of the intensity and frequency of the periodic electric
field. CDT has been theoretically analyzed for ac electric field driven
systems such as double quantum wells \cite{GrifoniM.1998,Grossmann1991,PhysRevB.66.085325},superlattices\cite{PhysRevB.47.6499},
double quantum dots (DQD's)\cite{Creffield2002,Brandes2004}, arrays
of QD's\cite{Creffield2010,Creffield2004}, nano-electromechanical
systems, as triple vibrating QD's\cite{Villavicencio2011}, strongly
correlated two dimensional QD's\cite{Creffield2002a} and Bose-Einstein
condensates\cite{Creffield2009,JGong2009}. Also recent experiments
on periodically driven cold atoms show super-Bloch oscillations which
present a strong dependence on the phase of the potencial driving\cite{PhysRevLett.104.200403}.

Spin qubits, consisting of two-level systems, can also be coherently
manipulated in DQD's. Electron spin resonance experiments\cite{Koppens2006,Nowack2007,Pioro-Ladriere2008}
measure coherent spin rotations of one single electron, a fundamental
ingredient for quantum operations. In the present work we propose
an additional way to control and manipulate spin qubits in double
quantum dots driven by ac periodic magnetic fields, by tuning the
intensity, frequency and relative phase of the ac magnetic fields
within each quantum dot. As we will see below, under particular field
configurations, charge localization within either the left or the
right dot can be induced. Furthermore, we will show how the ac magnetic
field can be tuned in order to lock the electron spin in its initial
spin state.

In the present work we analyze the effect of crossed dc and ac magnetic
fields on the electron dynamics in a double quantum dot. The static
field induces a spin splitting within each dot and the ac magnetic
field, which we consider linearly polarized, induces coherent spin
rotations. We will show that by tuning the parameters of the ac magnetic
fields: field intensity, frequency and phase, applied to the DQD,
coherent destruction of tunneling (and therefore charge localization)
can be achieved. We will show as well that ac magnetic fields allow,
in contrast with ac electric fields, to lock the electron spin in
the initial spin state at particular sets of ac field parameters.
How the system can be tuned to reach independently one or the other,
or both simultaneously will be discussed.\\
We consider Floquet theory\cite{Kohler.2005} to analyze the electron
dynamics in isolated DQD's, in the presence of crossed dc and ac magnetic
fields. We show how the different symmetries of the Hamiltonian determine
the conditions for inducing charge localization or dynamical spin
locking, and how they are related with the quasienergy spectrum. Our
results cover the case of different Zeeman splitting within each dot.
This configuration arises for instance, for vertical dots grown with
different materials (g-factor engineering\cite{Huang2010}) or due
to the inhomogeneous Overhauser field produced by Hyperfine interaction\cite{Vasyliunas1975,Dominguez2009,Lopez-Monis2011}.

\paragraph*{Model:}

\begin{figure}
\includegraphics[scale=0.25]{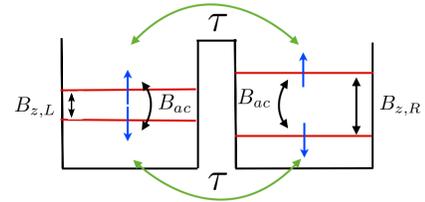}\caption{\label{fig:Double-quantum-dot}Double quantum dot with crossed ac
and dc magnetic fields: $\mathbf{B}\left(t\right)=\left(B_{ac}\left(t\right),0,B_{z}\right)$.
The Zeeman splitting in the left and right dot is given by $B_{z,L}$
and $B_{z,R}$ respectively. The parameter $\tau$ describes the tunneling
between the dots.}
\end{figure}

The system we consider is a double quantum dot with one electron (Fig.\ref{fig:Double-quantum-dot}),
subjected to crossed linearly polarized $B_{ac}\left(t\right)$, and
a $B_{z}$ magnetic fields. 

The Hamiltonian reads: 
\begin{equation}
H\left(t\right)=H_{0}+H_{\tau}+H_{dc}^{B}+H_{ac}^{B}\left(t\right)\label{eq:Hamiltonian}
\end{equation}

\begin{eqnarray*}
H_{0} & = & \sum_{i,\sigma}\epsilon_{g,i}c_{\sigma,i}^{\dagger}c_{\sigma,i}\\
H_{\tau} & = & -\tau\sum_{i\neq j}\left(c_{\sigma,i}^{\dagger}c_{\sigma,j}+h.c.\right)\\
H_{ac}^{B}\left(t\right) & = & \sum_{i}B_{ac,i}S_{x,i}\sin\left(\omega t+\phi_{i}\right)\\
H_{dc}^{B} & = & \sum_{i}B_{z,i}S_{z,i}
\end{eqnarray*}
 where the index $i=L,R$ (left/right), refers to the position in
the system, $\epsilon_{g,i}$ is the gate voltage in the $i$ dot,
$\sigma=\uparrow,\downarrow$ is the spin projection in the basis
of $S_{z}$ eigenstates, $\tau$ is the interdot tunneling parameter,
$\mathbf{S}_{i}=\left(S_{x,i},S_{y,i},S_{z,i}\right)$ is the spin
operator in second quantization at the $i$ position, and the intensities
of the ac and static magnetic fields within each dot are given by
$B_{ac,i}$ and $B_{z,i}$ respectively. Also, in difference with
ac electric fields, and due to the Hamiltonian symmetry in the present
case, the introduction of the phase parameter in the magnetic field
within each dot, $\phi_{i}$, will allow the definition of the generalized
parity symmetry (GP), i.e. a $\mathbb{Z}_{2}$ symmetry group different
of parity symmetry.

Previous works have shown that the GP symmetry, defined as: $\left\{ x\rightarrow-x,\ t\rightarrow t+T/2\right\} $,
is important for CDT\cite{Hanggi1993,GrifoniM.1998,Kierig2008}. This
parity operation defines a $\mathbb{Z}_{2}$ symmetry group, classifying
the solutions as even or odd. Our aim in this paper is to introduce
GP invariance in our system in order to localize and manipulate spin
qubits.

The Floquet theorem asserts that a general solution for a system with
$T$ periodicity can be written as $|\Psi_{\alpha}\left(x,t\right)\rangle=e^{-i\varepsilon_{\alpha}t/\hbar}|\phi_{\alpha}\left(x,t\right)\rangle$,
where $\varepsilon_{\alpha}$ is the quasienergy, and $|\phi_{\alpha}\rangle$
is the Floquet state with the property: $|\phi_{\alpha}\left(x,t\right)\rangle=|\phi_{\alpha}\left(x,t+T\right)\rangle$.
Because our results are obtained by means of Floquet theory, we can
consider GP symmetry as the natural extension of parity symmetry $\left\{ x\rightarrow-x\right\} $,
to the composed Hilbert space $H\otimes\mathcal{T}$ or Sambe space\cite{Sambe1973},
where $\mathcal{T}$ is the Hilbert space of all $T$-periodic functions.
The Floquet states are obtained by solving the Floquet Hamiltonian:
$\mathcal{H}\left(t\right)=H\left(t\right)-i\hbar\partial_{t}$, and
the scalar product is defined in Sambe space as: $\langle\langle\phi_{\alpha}\left(x,t\right)|\phi_{\beta}\left(x,t\right)\rangle\rangle=\frac{1}{T}\int_{0}^{T}\langle\phi_{\alpha}\left(x,t\right)|\phi_{\beta}\left(x,t\right)\rangle dt$.

Applying the GP operation to the Hamiltonian (Eq.\ref{eq:Hamiltonian}),
we obtain invariance for: $B_{z,L}=B_{z,R}$, $B_{ac,L}=B_{ac,R}$,
$\epsilon_{g,L}=\epsilon_{g,R}$ and $\phi=\phi_{R}-\phi_{L}=\pi$
(we consider $\epsilon_{g,L}=\epsilon_{g,R}$ from now on). We can
also define a $\mathbb{Z}_{2}$ invariance within a single dot: writing
the single dot Hamiltonian in the $S_{x}$ eigenstates basis, it is
clear that the system is invariant under \{$|\uparrow\rangle\leftrightarrow|\downarrow\rangle$,
$t\rightarrow t+T/2$\}, this is an internal $\mathbb{Z}_{2}$ symmetry
of the single dot Hamiltonian. We call this $\mathbb{Z}_{2}$ symmetry
\textit{generalized spin parity} (GSP).

If $\phi=0$ the Hamiltonian is parity invariant, but a change of
sign in all time dependent terms appears due to the addition of a
semiperiod, while the time independent terms remain with the same
sign, breaking the GP symmetry. Therefore, the difference of the phase
parameter determines if the symmetry of the system corresponds to
a parity invariant ($\phi=0$) or a GP invariant ($\phi=\pi$) Hamiltonian.
Approaches of quasienergies are frequently signatures of electron
localization via the suppression of tunneling. We will see below that
degeneracies of the quasienergies with opposite GP gives rise to CDT,
and degeneracies with opposite GSP result in spin locking, i.e., if
the electron is initially in some $S_{x}$ eigenstate, it will remain
locked in that particular spin projection.

\paragraph*{Results:}

The present configuration is not analytically solvable. We employ
numerical methods and perturbation theory in the Floquet formalism
in the high frequency limit \cite{GrifoniM.1998,Platero2004}, considering
the Zeeman splitting and the tunneling as the perturbation (up to
first order), both in the same footing. By high frequency limit we
mean that $\omega\gg\tau,B_{z}$. We obtain analytically the quasienergies
$\varepsilon_{\alpha}$, and numerically the time evolution of the
electronic states occupation.

\begin{figure}
\begin{centering}
\includegraphics[scale=0.6]{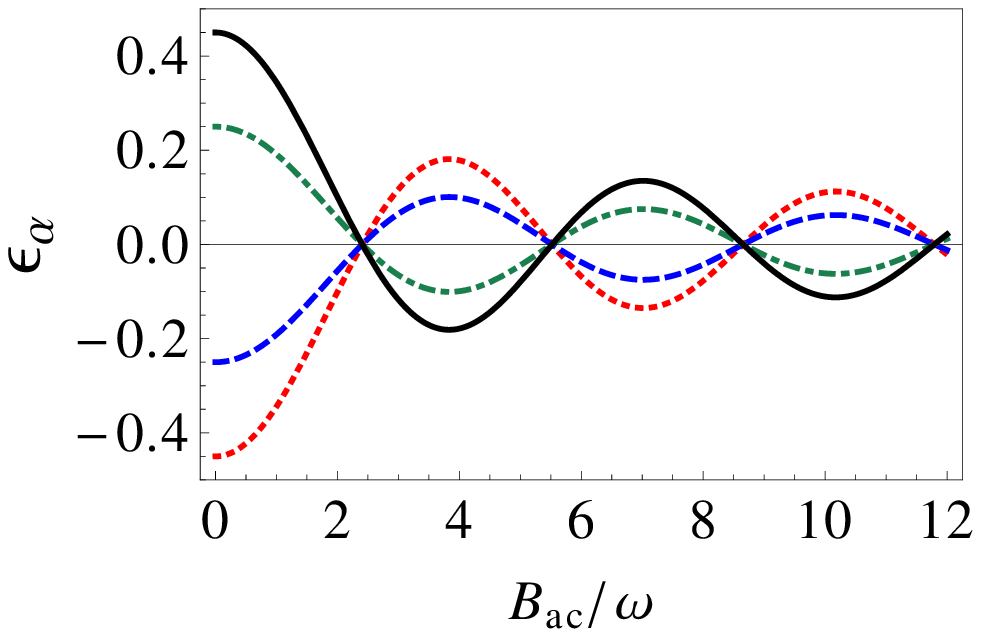} 
\par\end{centering}

\begin{centering}
\includegraphics[scale=0.6]{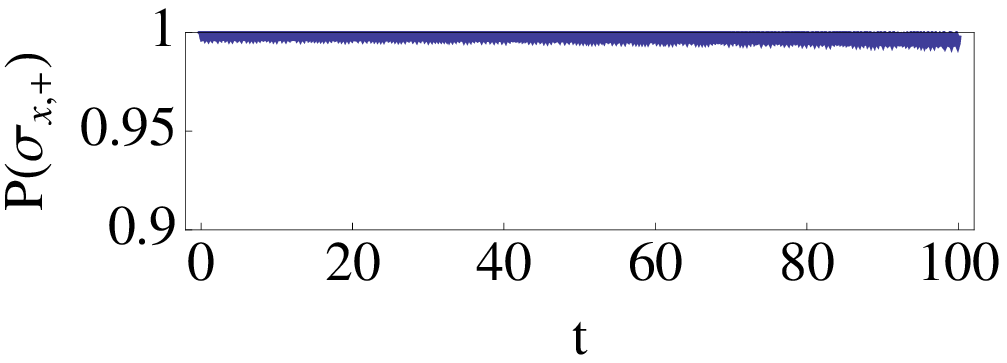} 
\par\end{centering}

\caption{\label{fig:Quasienergies1}(top) Quasienergies versus $B_{ac}/\omega$
for $\phi=\pi$. Note the crossings of all the quasienergies due to
the existence of two generalized parity symmetries $\mathbb{Z}_{2}$
(GP and GSP) for $J_{0}\left(B_{ac}/\omega\right)=0$. (bottom) Occupation
probability versus time for the initial state $|\Psi_{x,+}^{L}\rangle=\left(|\uparrow\rangle_{L}+|\downarrow\rangle_{L}\right)/\sqrt{2}$,
at the first crossing of the quasienergies ($B_{ac}/\omega\simeq2.404$).
The probability is almost constant in time, and the electron remains
in a well defined spin projection and spatially localized in the left
dot. In this case, both CDT (spatial localization) and dynamical spin
locking occurs at the first zero of $J_{0}\left(B_{ac}/\omega\right)$.
Parameters: $B_{z,L}=B_{z,R}=0.7$, $\omega=8$ and $\tau=0.1$ in
units $\mu_{B}=\hbar=1$. These parameters correspond to a $\omega=16\mu\text{eV}$,
$B_{z}=1.4\mu\text{eV}$ and $\tau=0.2\mu\text{eV}$.}
\end{figure}

Fig.\ref{fig:Quasienergies1}(top) shows the quasienergy spectrum
with symmetric Zeeman splittings ($B_{z,L}=B_{z,R}$), and $\phi=\pi$
within the first Brillouin zone, in the high frequency limit. The
parameters considered are of the order of those typical for transport
experiments in QDs (\cite{Pioro-Ladriere2008}).\\
 The quasienergies obtained by means of perturbation theory become:
\begin{equation}
\varepsilon_{s,s^{\prime}}=\frac{1}{2}\left(\pm B_{z}\pm2\tau\right)J_{0}\left(\frac{B_{ac}}{\omega}\right),\quad\left(s,s^{\prime}=\pm\right)\label{eq:quasi_energies0}
\end{equation}

\begin{figure}
\includegraphics[scale=0.6]{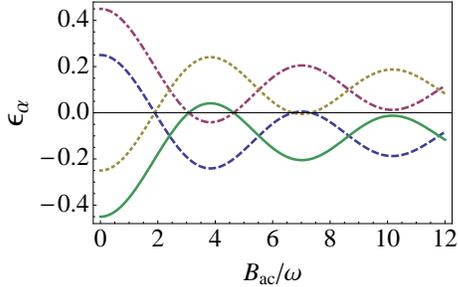} \caption{\label{fig:Quasienergies2}Quasienergies versus $B_{ac}/\omega$ for
$\phi=0$. GP symmetry is broken. Crossings between quasienergies
with opposite GSP appear, but the quasienergies with opposite parity,
do not cross due to a $2\tau$ splitting. Therefore, at the first
zero of $J_{0}$, CDT does not happen, and just spin locking is achieved.
$|\Psi_{x,+}^{L}\rangle$ is the initial state, being $|\Psi_{x,\pm}^{i=L,R}\rangle=\left(|\uparrow\rangle_{i}\pm|\downarrow\rangle_{i}\right)/\sqrt{2}$.
In this case, the electron performs Rabi oscillations between $|\Psi_{x,+}^{L}\rangle$
and $|\Psi_{x,+}^{R}\rangle$, remaining in the initial spin projection.
Parameters: $B_{z,L}=B_{z,R}=0.7$, $\omega=8$ and $\tau=1/10$.}
\end{figure}

The crossings of all the quasienergies in several points, according
to the Wigner-Von Neumann theorem\cite{Wigner1929}, are related to
quasienergies that belong to different symmetry groups. This fact
reflect the classification of the quasienergies in a $\mathbb{Z}_{2}\otimes\mathbb{Z}_{2}$
symmetry group (GSP and GP), where all the Floquet states are orthogonal
to each other. CDT and also dynamical spin locking has been found
in this configuration (Fig.\ref{fig:Quasienergies1}(bottom)) for
values of the frequency and intensity of the field such that $J_{0}\left(B_{ac}/\omega\right)=0$
(Eq.\ref{eq:quasi_energies0}). Crossings between quasienergies with
opposite GP result in charge localization in the left or right dot,
while crossings between quasienergies with opposite GSP result in
dynamical spin locking for the eigenstates of the $S_{x}$ matrix.
If we consider instead, that $\phi=0$, where the fields are in phase,
the quasienergy spectrum changes, disappearing the multiple crossings
(Fig.\ref{fig:Quasienergies2}). 

The measurement of the occupation probabilities shows that CDT does
not occur in this case, but dynamically induced spin locking is achieved.
These results indicate that CDT is strongly dependent of the spatial
inhomogeneity introduced by the difference of phase $\phi$, while
spin localization is not.

Now we consider an spatial inhomogeneity in the Zeeman splittings
(i.e. $B_{z,L}\neq B_{z,R}$), such that the parity and GP symmetries
are broken, resulting in avoided crossings between quasienergies with
opposite GP.

We have obtained, by means of perturbation theory, the quasienergies
for arbitrary Zeeman splittings ($B_{z,L}\neq B_{z,R}$) and their
dependence with $\phi$: 
\begin{eqnarray}
\varepsilon_{i,j} & = & \pm\frac{1}{2}\frac{B_{z,L}+B_{z,R}}{2}J_{0}\left(\frac{B_{ac}}{\omega}\right)\label{eq:quasienergies}\\
 &  & \pm\frac{1}{2}\sqrt{\Delta_{Z}^{2}J_{0}^{2}\left(\frac{B_{ac}}{\omega}\right)+\left(2\tau\right)^{2}J_{0}^{2}\left(\frac{B_{ac}}{\omega}\sin\left(\phi/2\right)\right)}.\nonumber 
\end{eqnarray}
 The indices $i,j=\pm,\pm$ label the parity, according to each symmetry
group ($i$ refers to the GP and $j$ refers to the GSP), and $\Delta_{Z}=\left(B_{z,L}-B_{z,R}\right)/2$.
A very good agreement between the analytical expression (\ref{eq:quasienergies})
and the numerical calculation has been found.\\

\begin{figure}
\includegraphics[scale=0.6]{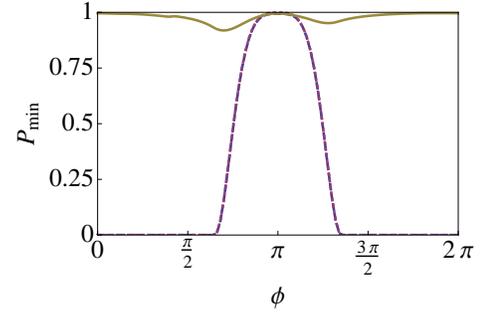}\caption{\label{fig:P_min}(Color online) $P_{min}^{k}$ versus the difference
of phase for the first zero of $J_{0}\left(B_{ac}/\omega\right)$,
i.e. $B_{ac}/\omega\simeq2.404$. Dashed line shows initial state
freezing ($P_{min}^{1}$, red), dotted line shows the spatial localization
($P_{min}^{2}$, blue), while the continuous line shows the spin locking
($P_{min}^{3}$, brown). In this case $P_{min}^{1}$ and $P_{min}^{2}$
are coincident. As the phase $\phi$ moves away from $\pi$, $P_{min}^{k=1,2}$
decreases, reducing and even destroying CDT. Spin localization ($P_{min}^{3}$)
holds for all $\phi$ (the state remains with a well defined spin
projection, but oscillating within the dots), because, although GP
symmetry is broken as $\phi$ varies from $\pi$, GSP remains. Parameters
considered: $\omega=8$, $B_{z,L}=B_{z,R}=0.7$, and $\tau=1/10$. }
\end{figure}

In order to analyze the $\phi$ dependence at fixed intensity, we
define the states $|\Psi_{x,\pm}^{L,R}\rangle=\left(|\uparrow\rangle_{L,R}\pm|\downarrow\rangle_{L,R}\right)/\sqrt{2}$
and the measure $P_{min}^{k}$, that takes the minimum value over
one hundred periods to be in either the initial state $|\Psi_{x,+}^{L}\rangle$
($k=1$), the left dot ($k=2$), or at the initial spin state projection,
delocalized between the left and right dot ($k=3$), when the system
time evolves from the initial state. Formally can be defined as $P_{min}^{k}=\text{min}\left\{ \sum_{i}\left|\langle\Psi_{k}^{i}|U\left(t,0\right)|\Psi_{x,+}^{L}\rangle\right|^{2},\ t\in\left[0,100T\right]\right\} $,
being $U\left(t,0\right)$ the time evolution operator obtained numerically
from the Floquet states and $i$ the number of states composing the
subspace we are measuring, ie. $k=1\rightarrow i=1$ being $|\Psi_{1}^{1}\rangle=|\Psi_{x,+}^{L}\rangle$,
$k=2\rightarrow i=1,2$ being $|\Psi_{2}^{1,2}\rangle=|\Psi_{x,\pm}^{L}\rangle$
and $k=3\rightarrow i=1,2$ being $|\Psi_{3}^{1,2}\rangle=|\Psi_{x,+}^{L,R}\rangle$.
Fig.\ref{fig:P_min} shows the measure $P_{min}^{k}$ as a function
of the phase difference between the ac magnetic fields within each
dot. 

\begin{figure}
\includegraphics[scale=0.6]{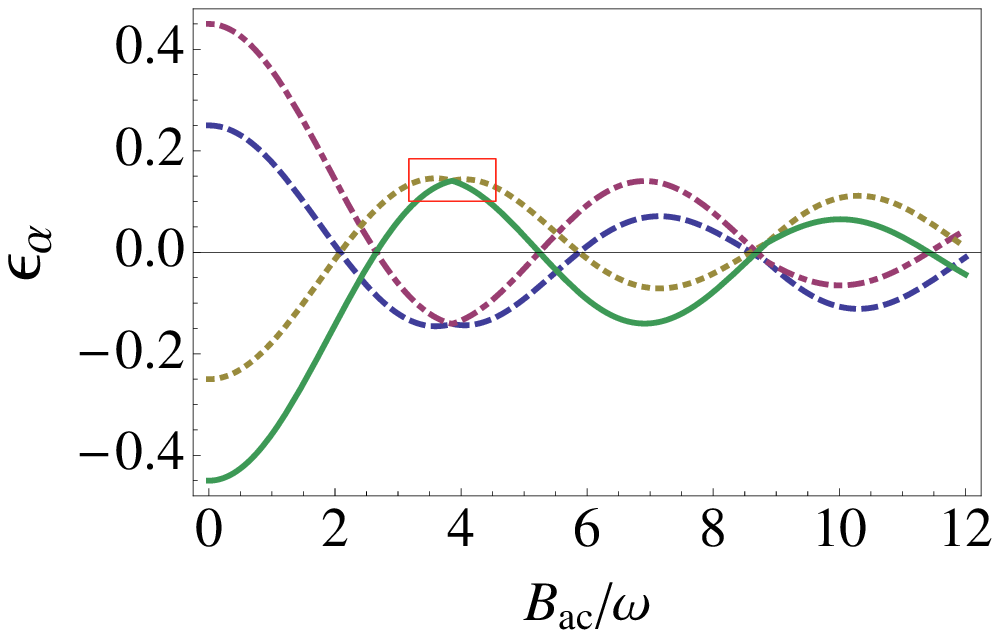}\includegraphics[scale=0.6]{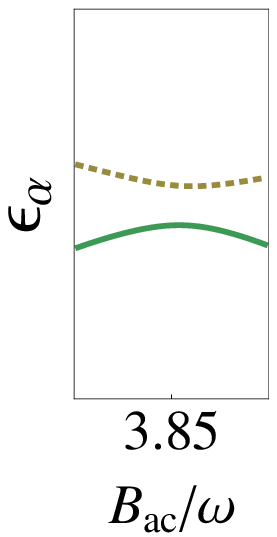}

\caption{\label{fig:Avoided}Quasienergies versus $B_{ac}/\omega$ for $\phi=3\pi/7$.
The avoided crossings arise from the lack of GP symmetry produced
by the difference of phase $\phi\neq\pi$. Right figure shows a zoom
of the framed area. Long time spatial localization is obtained in
such avoided crossings, due to the nearby degeneracy. At $B_{ac}\simeq5.5$
the crossing of quasienergies gives rise to dynamically induced spin
locking. Also for $B_{ac}\simeq8.8$ a nearby degeneracy between all
four quasienergies occurs, giving rise to charge and spin locking.
This is in good agreement with Eq.(\ref{eq:intensities}) for $n=3$,
where spatial and spin localization occur. $B_{z,L}=B_{z,R}=0.7$,
and $\tau=1/10$.}
\end{figure}

In Eq.(\ref{eq:quasienergies}) the Bessel functions have different
arguments due to the phase difference $\phi$. Therefore their zeros
will be shifted, matching just at some intensities, where $J_{0}\left(B_{ac}/\omega\right)=J_{0}\left(B_{ac}\sin\left(\phi/2\right)/\omega\right)=0$;
i.e. exactly at intensities where spin locking and charge localization
are recovered. In the high intensity limit (i.e. $B_{ac}\gg\omega\gg\tau,B_{z}$)
we find these values to be: 
\begin{equation}
\frac{B_{ac}}{\omega}=\frac{\pi\left(4n-1\right)}{4\sin\left(\phi/2\right)},\ \left(n\in\mathbb{Z},n>0\right).\label{eq:intensities}
\end{equation}
 Where we have used the asymptotic limit $J_{0}\left(x\right)\underset{x\gg1}{\sim}\sqrt{2/\left(\pi x\right)}\sin\left(x+\pi/4\right)$.
The case $n=3$ is shown in Fig.\ref{fig:Avoided} for $B_{ac}/\omega\simeq8.8$.

In case of $\phi\neq\pi$ and $\phi\neq0$, we can also consider avoided
crossings between quasienergies with opposite GP parity out of the
zeros of $J_{0}$ (inset in Fig.\ref{fig:Avoided}, shows $\varepsilon_{+,\pm}$
and $\varepsilon_{-,\pm}$). 

Fig.\ref{fig:Avoided} shows how just by tuning the intensity of the
ac field one can switch between the different regimes, i.e. spatial
localization regime (\textbf{$B_{ac}/\omega\simeq3.8$}), dynamical
spin locking ($B_{ac}/\omega\simeq5.5$) and both ($B_{ac}/\omega\simeq8.8$).
The value of the $P_{k}$ measured in the avoided crossing in Fig.\ref{fig:Avoided}
for $B_{ac}/\omega\simeq3.8$ are: $P_{min}^{2}=0.98$ and $P_{min}^{3}=0$,
confirming just spatial charge localization and spin rotations.

\paragraph*{Conclusions:}

We have shown that, in the high frequency regime, driving DQD's with
ac magnetic fields allows to achieve independently charge localization
and spin locking, by tuning the parameters of the ac magnetic fields.
In particular, modifying the intensity or the phase difference of
the ac magnetic fields applied to the quantum dots allows to select
spatial localization, spin locking or both regimes simultaneously
in the system, giving rise to a coherent control of the electronic
states. In fact, the change of frequency can be used to modify the
difference of phase of the ac magnetic fields between spatially separated
dots.

We explain the different localization regimes in terms of the parity,
GP or GSP symmetry in the system, which can be externally imposed
and modified. Our results can be extended to other systems consisting
on coupled two level systems with internal SU(2) and parity symmetry.

In summary, the combined localization effect of ac magnetic fields
on the spin and spatial degrees of freedom allows a novel way for
inducing, in a controlled way, charge localization and spin locking
in double quantum dots. It opens a new road for qubits manipulation,
a fundamental task for quantum information and quantum computation
purposes. These results are also interesting for potential spintronic
devices.
\begin{acknowledgments}
We acknowledge Dr. C.E. Creffield, Dr. S. Kohler and M. Busl for critical
reading of the manuscript and discussions. This work has been supported
by Grant No. MAT2008-02626 (MICINN) and by ITN under Grant No. 234970
(EU).One of us (A.G.L.) acknowledges the JAE Predoctoral program (MICINN). 
\end{acknowledgments}
 \bibliographystyle{phaip}
\bibliography{Paper_bib}

\end{document}